# A Conversation with Monroe Sirken

**Barry I. Graubard, Paul S. Levy and Gordon B. Willis**

*Abstract.* Born January 11, 1921 in New York City, Monroe Sirken grew up in a suburb of Pasadena, California. He earned B.A. and M.A. degrees in sociology at UCLA in 1946 and 1947, and a Ph.D. in 1950 in sociology with a minor in mathematics at the University of Washington in 1950 where Professor Z. W. Birnbaum was his mentor and thesis advisor. As a Post-Doctoral Fellow of the Social Science Research Council, Monroe spent 1950–1951 at the Statistics Laboratory, University of California at Berkeley and the Office of the Assistant Director for Research, U.S. Bureau of the Census in Suitland, Maryland.

Monroe visited the Census Bureau at a time of great change in the use of sampling and survey methods, and decided to remain. He began his government career there in 1951 as a mathematical statistician, and moved to the National Office of Vital Statistics (NOVS) in 1953 where he was an actuarial mathematician and a mathematical statistician. He has held a variety of research and administrative positions at the National Center for Health Statistics (NCHS) and he was the Associate Director, Research and Methodology and the Director, Office of Research and Methodology until 1996 when he became a senior research scientist, the title he currently holds.

Aside from administrative responsibilities, Monroe's major professional interests have been conducting and fostering survey and statistical research responsive to the needs of federal statistics. His interest in the design of rare and sensitive population surveys led to the development of network sampling which improves precision by linking multiple selection units to the same observation units. His interest in fostering research on the cognitive aspects of survey methods led to the establishment of permanent questionnaire design research laboratories, first at NCHS and later at other federal statistical agencies here and abroad.

Monroe has been active in serving the statistical community. He has served on many committees of the American Statistical Association (ASA) and the Washington Statistical Society (WSS). He is a charter member of the Federal Committee on Statistical Methodology (FCSM) and chairs its research subcommittee that oversees a grants program in statistical and survey research that is funded by a consortium of federal statistical agencies, and administered by the National Science Foundation. He is a Fellow of the American Statistical Association and the American Association for the Advancement of Science, and is an elected member of the International Statistical Institute. He is the recipient of the Public Health Service Superior Service Award, and the ASA WSS Roger Herriot Award for Innovation in Government Statistics.

*Barry I. Graubard is Senior Investigator, Biostatistics Branch, Division of Cancer Epidemiology and Genetics, National Cancer Institute, 6120 Executive Blvd, Bethesda, Maryland 20892, USA e-mail: graubarb@mail.nih.gov. Paul S. Levy is RTI Senior Fellow, Statistical Methods in Health Sciences, and Senior Research Statistician, RTI International, 3040 Cornwallis Rd. PO Box 12194, Research Triangle Park, North Carolina 27709-2194, USA e-mail: levy@rti.org. Gordon B. Willis is Cognitive Psychologist, Applied Research Program, Division of Cancer Control and Population Sciences, National Cancer Institute, 6130 Executive Blvd, Bethesda, Maryland 20892, USA e-mail: willisg@mail.nih.gov.*

This conversation ranges over Monroe's education and research agendas during a 55-year career as a government statistician. The conversation took place at the National Center for Health Statistics, Hyattsville, Maryland, in three sessions during the spring of 2006.

### EARLY LIFE

**Graubard:** Could we begin by talking about your early life?

**Sirken:** My mother was born in Scranton, Pennsylvania and married my father in 1919. My father was born in Poland and his family migrated to the United States about 1910 when he was about 12 years old. He was a disabled World War I veteran and died when I was about 14 years old. I was born in New York City, moved to upstate New York when I was about two, and ten years later moved to California. My mother, sister and I lived in Sierra Madre, a suburb of Pasadena, where I attended school before moving to Los Angeles and graduating from Fairfax High School in June 1938.

### UNIVERSITY OF CALIFORNIA AT LOS ANGELES

**Graubard:** Then you went to the University of California, Los Angeles. Why that school?

**Sirken:** Well, I never considered going elsewhere. UCLA was virtually free for California high school graduates with good grades. As I recall, UCLA tuition my first semester in September 1938 was $29 plus $4 for a student membership card that entitled me to admission for all UCLA sports events. Another financially related reason is, like most students attending UCLA at that time, I couldn't afford to live on campus, and UCLA was close enough to where we lived that I could commute. Foremost, I thought UCLA was a great university.

**Graubard:** I believe that your B.A. and M.A. were in the social sciences, and I wonder how you became interested in statistics and mathematics?

**Sirken:** I got a B.A. in sociology in 1946 and the following year an M.A. in anthropology and sociology. How I became interested in statistics is a longer story. I began UCLA as a pre-med major with intentions of going to medical school but during my sophomore year, I contracted tuberculosis. After recovering my health, about three years later, I returned to UCLA in 1943. In my financial situation, it was unrealistic to think of medical school. So I changed my major to sociology thinking that I might become a social worker. However, some faculty in the sociology department encouraged me to think about becoming a sociologist and advised me to take as much mathematics as possible. So I did, without really appreciating how math would help me as a sociologist or anything else. About 1945, I took my first course in mathematical statistics from Dr. Paul Hoel.

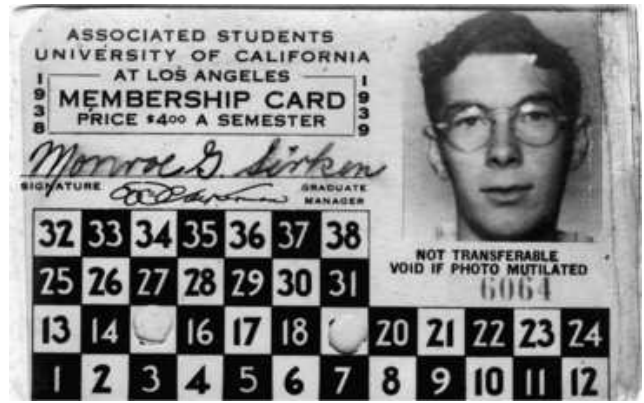

Fig. 1. *Monroe's UCLA Student Membership Card, 1938.*

**Graubard:** Why didn't you stay at UCLA for your Ph.D.?

**Sirken:** Well, for one thing, students were not encouraged to get the Ph.D. at the same universities at which they were undergraduates. However, the more important reason is that I had become quite interested in quantitative sociology, and UCLA did not offer that kind of graduate program at the time. I was awarded a fellowship in the sociology department at the University of Washington, which my UCLA advisors said was strongly quantitatively oriented.

### UNIVERSITY OF WASHINGTON

**Graubard:** So what department did you actually end up in at the University of Washington?

**Sirken:** I started out as a teaching assistant in the sociology department. But one of the greatest strokes of good fortune in my professional life occurred soon after I arrived in Seattle though I didn't realize it at the time. And that good fortune was ultimately responsible for me ending up straddling two departments.

In 1947, the math department at the University of Washington began to offer a two-year sequence of graduate courses in statistics and the person in charge of the program was Professor Z. W. Birnbaum. I was unsure that I qualified so I went over to Dr. (Bill) Birnbaum and described my background in math and statistics and my interests in quantitative sociology. To make a long story short he accepted me, and I took the courses he offered in 1947–1948 and did very well. He was an excellent instructor, took an interest in his students and they very much appreciated him. I recall that on the final day of the last class in the sequence, the students presented Bill with a goldplated multicolor pen with the inscription "nature is not vicious" because that was the phrase he often used when the math got really complicated. When Bill died a few years ago, his daughter told me that she found the inscribed pen on top of his desk.

In 1948, Dr. Birnbaum offered me a job as a research assistant in his newly established statistics laboratory, which I instantly accepted. However, I had already accepted a position as a research assistant in the newly formed Washington Public Opinion Poll (WPOL), which was housed in the sociology department, and where I was becoming acquainted with sampling and sample survey methods. So during 1948, I had distinct appointments in the sociology and mathematics departments which I believe was quite unusual at the time and may have been illegal.

**Graubard:** I suppose that kind of interdisciplinary type work would be encouraged now; back then it must have been pretty unusual.

**Sirken:** Yes, I believe it was quite unusual in those days. I was just plain lucky to be in the right place at the right time. It happened because George Lundberg, Chair of Sociology Department, encouraged interdisciplinary research at the intersection of the social and mathematical sciences, and Z. W. Birnbaum was willing to take a chance on a social scientist.

**Graubard:** Professor Birnbaum is well known for his theoretical contributions in mathematics and statistics. It's unusual, I would think, for someone with your background to work at his level.

**Sirken:** I could not and did not work with Bill on those kinds of problems. He helped me solve my applied statistical problems, and my problem then was measuring "bias due to nonresponse in sample surveys." It so happened that my survey research problem was nested in Bill's broader interest in truncated population distributions. We successfully derived the expression the mean squared error that minimized the joint effects of sampling and nonresponse bias in survey estimates of categorical variables (Birnbaum and Sirken, 1950).

I worked in Bill's laboratory for two years, and wrote my thesis under his direction. Bill taught me how to think in terms of statistical models. That was an invaluable gift.

**Graubard:** What did you do after completing your Ph.D.?

**Sirken:** Well, I got the Ph.D. in June, 1950 and stayed in Bill's laboratory that summer working on the problem of optimizing the scheduling of callbacks of survey interviewers. I got this idea while writing the Ph.D. thesis on bias due to nonresponse in sample surveys. I learned much later that it was a linear programming problem.

## POST-DOCTORAL FELLOWSHIP AND FIRST JOB

**Graubard:** What was your first position after getting the Ph.D.?

**Sirken:** In the fall of 1950, I headed to the Statistics Laboratory at the University of California, Berkeley. I had a Social Science Research Council Post-Doctoral Fellowship and was planning to spend most of the next 12 months at Berkeley. I took courses that fall and the following spring from Jerzy Neyman and Erich Lehman and consulted while there with Ed Barankin. Ed thought he might have the solution to my linear programming problem, which I had originally solved for relatively simple situations, but I was unsuccessful in applying his theory. I left Berkeley in June, 1951.

**Graubard:** So that was a case when theory didn't solve a practical problem for you. After working at Berkeley for eight or nine months, what did you do?

**Sirken:** I was pretty sure that I wanted to learn more about sampling and survey research and what better place to get that kind of experience than at the U.S. Census Bureau. So I continued my fellowship there.

**Graubard:** Whom did you work with at the Census Bureau?

**Sirken:** Well, I was located in Morris Hansen's office. The staff included really outstanding people including Bill Hurwitz, Joe Daly, Max Bershad and Margaret Gurney. These people were part of the central staff and there were others also technically responsible to Morris who worked throughout the Bureau including Joe Steinberg, Joe Waksberg and Harold Nisselson, to name a few. However,

I was mostly involved with a small group of survey methodologists including Eli Marks and Leon Pritzker who were working on the Post-Enumeration Survey (PES). The PES evaluated the quality of the 1950 Census of Population by renumerating a sample of the census population. By the time I arrived, the PES fieldwork had been completed and the staff was programming the UNIVAC to run the PES tabulations. I was very impressed with the research at the Census Bureau so after the post-doctoral expired I took a job with Morris Hansen's group. I got involved in several projects. I particularly remember working with Hal Nisselson and Ted Woolsey on the pretest of the National Health Survey in San Jose, California. Little did I realize then how relevant that experience would become in my future work.

**Graubard:** So was your academic career over?

**Sirken:** I didn't think so at the time. When I arrived at the Census Bureau in 1951, my original plan was to stay three months and then take an academic appointment. In fact, I have been in government for over 55 years with brief visiting appointments in biostatistics departments at University of California at Berkeley and University of North Carolina at Chapel Hill.

## NATIONAL OFFICE OF VITAL STATISTICS

**Graubard:** What made you leave the Census Bureau and work at the National Office of Vital Statistics?

**Sirken:** It wasn't my decision. There was a so-called Reduction in Force (RIF) throughout government soon after Dwight Eisenhower moved into the White House in 1952. I didn't have tenure at the Census Bureau, and was RIF'ed in midyear 1953. A couple of months later, Morris Hansen got me a job at the National Office of Vital Statistics (NOVS) which was quite a feat in view of the government-wide employment freeze on outside hires. Well, Morris Hansen and Dr. Halbert Dunn, Director of NOVS, were old friends and Morris apparently convinced Dr. Dunn that I was a very competent actuarial mathematician. Later, when Dr. Dunn interviewed me, I noted my very limited knowledge (virtually none) in actuarial science, and luckily he thought I was being modest.

**Graubard:** So you launched into an area that you had no real experience.

**Sirken:** That's right. However, I had a couple of fine mentors. T. N. E. Greville, an outstanding actuarial mathematician and statistician who had been my predecessor at NOVS, was an exceptionally good communicator and his publications were very helpful. Also, Mortimer Spiegelman, a well-known demographer at Metropolitan Life Insurance, was my collaborator in constructing the 1950 U.S. and state life tables.

**Graubard:** What did you do after you finished the life table assignment, and who did you work with?

**Sirken:** Unlike the U.S. Census Bureau where most national population data are collected periodically in censuses, at NOVS, national vital statistics (births, deaths, marriages and divorces) are compiled as by-products of information reported on vital records. Because vital records serve primarily as legal documents, the information reported on vital records is necessarily limited and virtually changeless. I had an understanding with Dr. Dunn that after completing the life tables I would work on sample survey methods to improve vital statistics. In 1955, just after the life table project was completed, I had a chance to do just that when Bill Haenszel, a well-known epidemiologist at the National Cancer Institute (NCI), proposed a collaborative research project in which NOVS would design and test sample survey methodologies to collect retrospective residence and smoking histories for samples of deceased persons from their surviving relatives. With funding support from NCI, a small statistical unit, including Mort Brown and Jim Pifer and myself and a clerical staff, was established in NOVS to conduct Haenszel's pilot study. Soon after the successful completion of that pilot, NOVS established a long-range research sample survey program to expand the scope and improve the quality of vital statistics by conducting retrospective sample surveys linked to birth and death records (Sirken, 1963).

**Graubard:** So you were using the idea that people like Morris Hanson had promoted at the Census Bureau that by sampling you could expand the scope and improve the quality of vital statistics.

**Sirken:** The NOVS survey program was sustained by conducting work for other government agencies. For example, Haenszel expanded the lung cancer pilot study into a national mortality survey, and he arranged with the Census Bureau to collect information on smoking and residence histories for the national population in the Current Population Survey (CPS). Thus, we were able to estimate national lung cancer death rates by smoking habits and residence histories (Haenszel, Loveland and Sirken, 1962). As I recall, these findings were cited in the first report of the Surgeon General on Smoking and Health.

**Graubard:** And actually that's one of the first examples of a population-based case-control study, where the mortality follow-back survey provided exposure and other covariate information for the lung cancer cases and the CPS provided these variables for the control sample of the population at risk.

**Sirken:** Exactly. I have always felt that the linked mortality/population sample survey methodology deserves much more attention than it has received from epidemiologists.

**Graubard:** Weren't you also involved in designing other surveys for federal health agencies?

**Sirken:** Yes. NOVS developed the methodology of the follow-back surveys linked to birth records, and with funding from the Division of Radiological Health, U.S. Public Health Service, conducted the first national natality survey on the exposure of pregnant women to medical radiation. With funding from the U.S. Public Health Service, NOVS contracted with the Census Bureau for a CPS supplement on the population's utilization of the Salk vaccine. Using data from the CPS polio supplement, NOVS produced the first national statistics on the utilization and effectiveness of the Salk vaccine (Sirken, 1962). Thereafter, the Public Health Service often used the CPS supplements to monitor the immunization status of the national population. On another occasion, the U.S. Children's Bureau asked NOVS to conduct a survey on the prevalence of cystic fibrosis, a debilitating and often lethal pediatric disease, and to do so within something like a hundred days in order to comply with a Congressional request. An unexpected estimation problem in that medical provider survey ultimately led to the development of a new kind of sampling called network sampling. These were very exciting days when the findings of the NOVS sample surveys were used in real time to address important public health problems.

**Graubard:** Much of your own early research and research collaborations involved measuring the quality of survey data. How did this come about?

**Sirken:** Well, it started with my thesis and continued with my work at the Census Bureau. Also, my training in demography was a factor. I tended to think about the demographic aspects of survey measurement errors, and in the back of my mind I wondered how information about the population being surveyed could be used in the survey measurement process to reduce sampling and measurement errors. This way of thinking helped me to recognize and exploit research opportunities that I might otherwise have overlooked.

**Graubard:** That way of thinking explains how you got involved in research on network sampling and the cognitive aspects of survey methods. Both areas seek to improve the survey quality by using information about the population being surveyed. It seems to me that you took almost a personal responsibility for the total quality of your surveys—sampling and nonsampling errors.

**Sirken:** I guess so. Generally, NOVS was not responsible for survey data analyses. That was the responsibility of the contracting agency. But NOVS was responsible for describing the data limitations, both sampling and nonsampling errors, and we took those responsibilities very seriously.

## EARLY DAYS AT THE NATIONAL CENTER FOR HEALTH STATISTICS

**Graubard:** When and why did you leave NOVS?

**Sirken:** In 1960, NOVS was merged with the National Health Survey (NHS) to form the National Center for Health Statistics (NCHS). Without any action on my part, I became a charter member of the NCHS, though for a couple of years afterward I remained in NOVS, which was renamed the Division of Vital Statistics (DVS). The NHS was the other NCHS division. The NHS, located in the Office of the Surgeon General, was chartered in 1956 and was responsible for the National Health Interview Survey (NHIS)—a large national face-to-face household survey that was and is still fielded by the Census Bureau. You may recall that at the Bureau in 1952, I had worked on the NHIS San Jose pretest, and next year NHIS will be celebrating its 50th anniversary.

The NCHS is the official federal agency responsible for producing national health statistics, and it is noteworthy that the Congressional Act that created the NCHS empowered NCHS to develop the survey methodology most suitable for collecting health statistics. A lot of credit for establishing the NCHS goes to Ted Woolsey whom I believe helped draft the enabling legislation, and he also directed the development of the NHIS. Forrest Linder was the first NCHS Director. He was a demographer who had formerly worked in the Vital Statistics Division at the Census Bureau and had recently resigned from the United Nations to become the NHS Director. I knew Forrest Linder and the other NHS staff including Ted Woolsey, Ozzie Sagen, Phil Lawrence,

Walt Simmons and Earl Bryant because NHS and NOVS occupied quarters in the same building and the staffs often met for lunch.

**Graubard:** You've given us an overview of the formation of the National Center for Health Statistics and noted some of the staff. What kind of place was NCHS in those early days?

**Sirken:** A very exciting place. Forrest Linder was a good administrator and had the vision of developing a family of national data systems that intersected all important health-related activities of the population that would be capable of meeting the increasing needs for national health and vital statistics. I was particularly excited because fulfilling Forrest's vision implied the need for a strong survey methods research program.

Even before the NCHS was established, the NHS had begun to conduct pilot studies of the National Health Examination Survey, now the National Health and Nutrition Examination Survey (NHANES)—a medical examination survey that uses mobile medical trailers to physically examine random samples of the national population. When the NCHS was established, the vital record follow-back surveys became an integral part of NCHS data systems. Soon after NCHS was established, we began to think about developing a family of national health care provider surveys of hospitals, physicians, clinics, home health services, etc.

As a result of an NCHS reorganization in 1963, there were now four divisions including the Division of Health Records Statistics (DHRS). The DHRS was responsible for health care provider surveys and vital record linked surveys. I became the first DHRS Director.

**Graubard:** What were the surveys that eventually came out of DHRS?

**Sirken:** During my four years at DHRS, we established the Master Facility Inventory of Health Care Providers, the National Hospital Discharge Survey, the National Nursing Home Survey and the Linked Birth Record Sample Survey.

Just before Dr. Linder retired in 1967, he appointed me the director of the newly created Office of Statistical Methods with responsibility for directing the NCHS survey and statistical methodology programs. Earl Bryant was the deputy director. It was a wonderful job and offered many opportunities. Our work was divided between serving as statistical consultants and advisors to the NCHS programs and conducting research relevant to the mission of NCHS that emerged from the consulting activities. We hired very talented people like Bob Casady, Paul Levy, Iris Shimizu and you, and over the years we welcomed input from many outstanding statisticians including Jerzy Neyman, Z. W. Birnbaum, Gad Nathan, Richard Royal, T. N. E. Greville, Phil McCarthy, Chin Long Chiang and Tom Jabine. Their reports were published in the NCHS Series 2 reports on Data Evaluation and Methods Research.

**Graubard:** I'm curious about how you were able to establish these research collaborations and what kind of support did you get from NCHS.

**Sirken:** Of course, Center support was essential. Relationships with scientists in academia were actively encouraged and supported by NCHS. How-ever, I think the key to attracting outstanding people is to have interesting problems for them to work on.

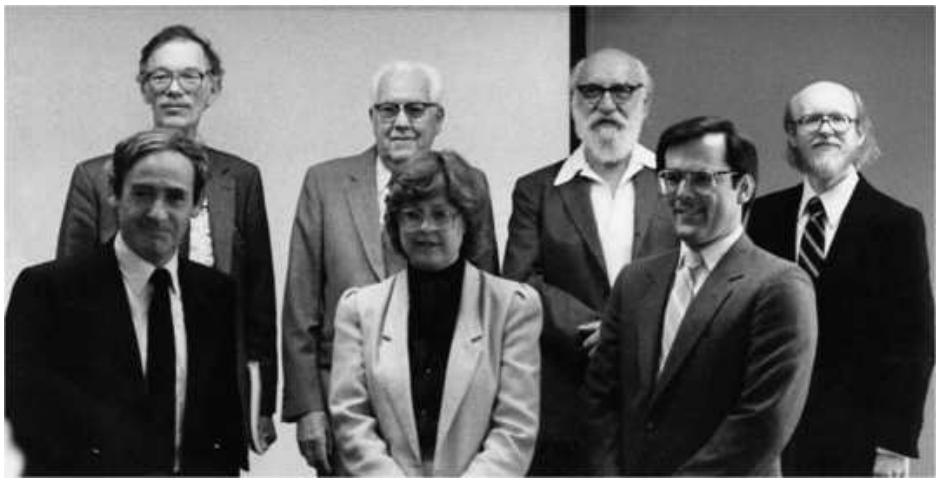

FIG. 2. *NCHS Seminar 1984. Front row l. to r.: Gad Nathan, Judy Lessler, Keith Eberhardt. Second row l. to r.: Monroe Sirken, Morris Hansen, Ben Tepping, Bob Fay.*

**Levy:** During your many years at NCHS you've led a large number of projects. What have been your major research interests and which projects do you regard as your personal favorites?

**Sirken:** Methodological research conducted by federal statistical agencies is severely constrained by short-term programmatic needs. Resources are hard to come by to pursue many interesting problems that emerge as by-products of short-term-oriented research. Nevertheless, by tapping a variety of sources I was able to obtain funds to research some interesting problems myself and with collaborators, and to foster methodological research relevant to long-term needs of federal statistical agencies. My favorite research projects are network sampling and integrated sample survey design. My main efforts to foster research involved the CASM movement, which fostered research on the cognitive aspects of survey methodology, and a grants program that supports investigator-initiated research projects that are oriented to the future needs of federal statistical agencies.

## NETWORK SAMPLING

**Levy:** Let's begin with network sampling. What is it and how does it differ from conventional sampling?

**Sirken:** The essential difference between network and conventional sampling is the kind of counting rule used to link the elements of the target population to the selection units at which they are eligible to be counted (Sirken, 1997). Conventional sampling depends on unitary counting rules that uniquely link each population element to one and only one selection unit. Network sampling depends on multiplicity counting rules that do not limit the number of selection units that are linked to the same population elements. For example, conventional sampling applies in household sample surveys that use the de jure residence rule that uniquely links each individual to his/her usual place of residence. On the other hand, multiplicity sampling would apply in household sample surveys that use the counting rule that links individuals to their de jure and de facto residences.

**Levy:** The solution of the estimation problem in the New England Cystic Fibrosis Survey ultimately led to the development of network sampling. How did that happen?

**Sirken:** In the New England Survey, the estimate of cystic fibrosis prevalence was based on a sample survey in which medical providers reported their cystic fibrosis patients. We did not appreciate until after the survey was conducted that most cystic fibrosis patients were treated by and eligible to be reported by multiple medical sources. Hence, the conventional sampling estimate of cystic fibrosis prevalence which implicitly assumes each patient is linked to one and only one medical source would have been biased. After I got Bill Birnbaum interested in this estimation problem, we developed three unbiased estimators of disease prevalence in stratified sample surveys of medical providers with overlapping patient case loads (Birnbaum and Sirken, 1965). The estimators utilize information about the multiplicity of linkages between patients reported in the sample survey and their medical providers. The simplest of the estimators, the so-called multiplicity estimator, weights each patient report by the inverse of the patient's multiplicity, namely, the inverse of the number of different providers that treated the patient.

**Levy:** I remember in your original paper with Dr. Birnbaum, you referred to the unbiased estimation technique as multiplicity sampling or multiplicity estimation. When did you actually begin using the term network sampling?

**Sirken:** Initially, our estimation technique was applied when surveys inadvertently used multiplicity rules and then it was called multiplicity estimation or multiplicity sampling. After demonstrating that multiplicity sampling is potentially more efficient than conventional sampling, I realized that our estimation technique also applied when multiplicity rules were deliberately used to improve sample survey efficiency (Sirken, 1970) especially in surveys of rare and elusive populations (Sudman, Sirken and Cowan, 1988). I coined the expression network sampling because the precision gains of using multiplicity instead of unitary rules depend on the statistical properties of the networks formed by the linkages between population elements and selection units. For example, assume the rare disease prevalence survey is based on a simple random sample of persons. In conventional sampling, each person responds for himself or herself. Suppose in network sampling, in addition to each person responding for himself, the person's siblings respond for him. Assuming all patients have the same number of siblings, $k$, and at most one sibling has the disease, the ratio of the sampling errors of network and conventional sampling is approximately the inverse of $k$.

Network sampling also has potential to improve the accuracy of survey estimates of household surveys of sensitive populations where the reports of close friends and relatives may be more accurate than those of the persons themselves, and in household surveys of elusive populations that are not linked to single places of residence. However, care must be taken in selecting the "right" counting rule because the rule that increases precision often decreases accuracy.

**Levy:** Is that what you meant by the "counting rule strategy"?

**Sirken:** Exactly. The counting rule strategy involves selecting the counting rule (unitary or multiplicity), which together with the other survey design features minimizes the mean square error of the survey estimate for fixed costs. Selecting the optimum counting rules requires a good demographic and sociologic knowledge of the target populations.

**Levy:** What kinds of survey experiments did you do to test the counting rule strategy, and who were your collaborators?

**Sirken:** The experiments compared the sampling and nonsampling errors of conventional and network sampling in household sample surveys of rare and sensitive populations—the kinds of surveys that often challenge conventional sampling. The rare population experiments were cancer and diabetes prevalence, and birth, marriage and death incidence. The sensitive issue experiments were alcohol and illicit drug use. In these experiments, the de jure residence rule was compared with multiplicity counting rules that incorporate the de jure rule. In the alcohol and illicit drug use surveys, for example, the multiplicity rules were based on self and friendship linkages, and in the rare population surveys, multiplicity rules were based on self and kinship linkages. For example, Trish Royston and I worked on the design effects counting rules in mortality surveys (Royston and Sirken, 1978).

**Levy:** Didn't one experiment involve an NCHS data system?

**Sirken:** Yes, an experiment was embedded in an NHIS supplement on diabetes which used multiplicity rules based on self and kinship relationships with parents, children and siblings. NHIS respondents reported whether they or close relatives had diabetes, and then reported the multiplicities of each reported diabetic. As I recall, diabetes prevalence estimates based on kinship rules were substantially larger and considerably closer to independent sources of diabetes prevalence estimates, and the network sampling errors were about half as large as the conventional sampling errors. Barry Graubard and I worked on this project.

**Levy:** As I recall, you and your collaborators were also busy expanding network sampling theory.

**Sirken:** Because network sampling does not specify rules for selecting samples, technically speaking it is not a sampling technique—it is an estimation technique. In our original paper, Bill Birnbaum and I developed three unbiased estimators for stratified network sample surveys. Much of our work in expanding network sampling theory involved applying network sampling in more complex and novel sample survey designs, and deriving the unbiased estimators and variances. For example, you and I collaborated in deriving the unbiased estimator and variance of ratios of random variables in stratified network sample surveys (Sirken and Levy, 1974), and Bob Casady and I collaborated with Gad Nathan of Hebrew University on the network sampling designs of dual system surveys (Casady, Nathan and Sirken, 1985). Also, I did some work on the components of variance of multiplicity estimators (Sirken, 1972).

## INTEGRATED SURVEY DESIGN

**Levy:** Network sampling research at NCHS was interrupted in the early 1980s. What happened?

**Sirken:** NCHS was quite a different place in the late 1970s and early 1980s than when established in 1960, and new kinds of research problems were emerging. I was Associate Director of Research and Methodology and the Director of the Office of Research and Methodology and Jim Massey, Andy White and Bob Casady succeeded by Randy Curtin were section chiefs.

By the early 1980s, the three major NCHS health surveys, the NHIS, the health care provider surveys, and the National Health and Nutrition Examination Survey (NHANES) were well established. Additionally, two new population sample surveys had been successfully fielded—the National Medical Expenditure Survey (NMES) and the National Survey of Family Growth Survey (NSFG). Some surveys, such as the NHIS, were continuous and others, like the NMES and the NSFG, were conducted periodically. Each survey had been independently designed and was being independently implemented. Dorothy Rice, then NCHS Director, established a Periodicity Committee of the Center's senior staff to recommend ways to reschedule the periodicity of conducting the Center's data systems in order to be able to maintain existing programs and undertake new programs. The Periodicity report compared the consequences of several short-term periodicity plans, and mentioned a long-term research plan which envisioned the possibility of integrating the sample designs of NCHS's independently designed population sample surveys.

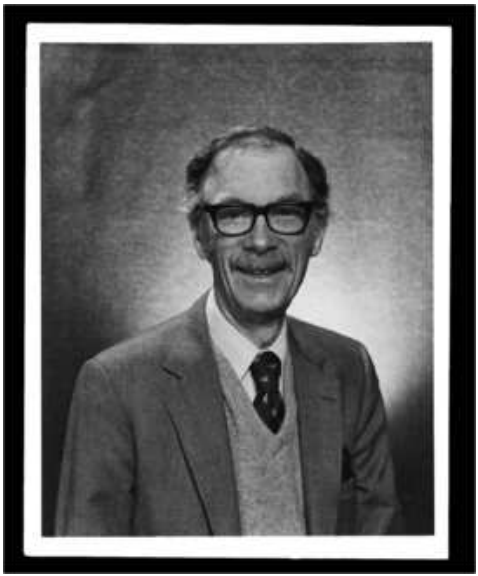

Fig. 3. *Monroe Sirken as NCHS's Associate Director of Research and Methodology, 1980.*

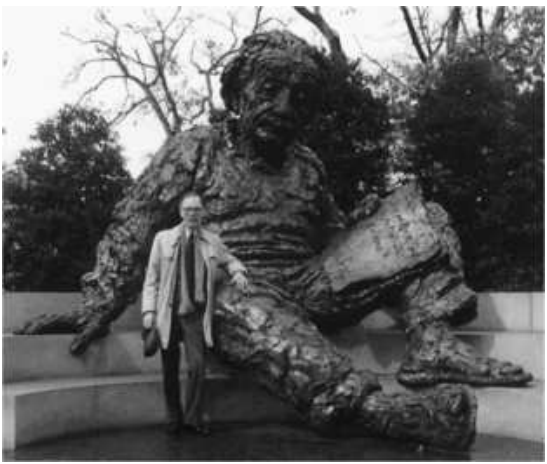

Fig. 4. *Dr. Albert Einstein and admirer, 1983.*

**Levy:** How did you intend to integrate NCHS's data systems and what made you think it would work?

**Sirken:** We proposed to use the NHIS listings of households and persons as the master sampling frame for NCHS's other population-based sample surveys including the NHANES, NMES and NSFG (Sirken and Greenberg, 1983). Subsequently, the proposal was expanded, I believe at Iris Shimizu's suggestion, to nest the health care provider surveys in the NHIS primary sampling units (psu's). There were several reasons why it made sense to use the NHIS as the hub of the integrated survey design system. With almost 50,000 households interviewed annually, NHIS was by far the largest of the NCHS population sample surveys and provided opportunities to oversample relatively small population domains, like race and ethnicity. Consistently high response rates, greater than 90%, assured that NHIS sampling frames would be relatively complete. Also, the NHIS collects a vast amount of health and related data about individuals and households that are relevant to the analytic objectives of NCHS's other population surveys.

**Levy:** So what happened next?

**Sirken:** There were confidentiality and funding problems. The NHIS is conducted for NCHS by the U.S. Bureau of the Census. Following each Decennial Population Census, the Census Bureau selects the NHIS household samples for the next decade. The penultimate sampling stage of that sampling process involves using the listings of households enumerated in the Decennial Population Census and that posed a serious administrative problem for the proposed integrated survey design because Census household listings are considered confidential Census data. Hence, any surveys that might in the future be linked to the NHIS would have to be conducted by the Census Bureau because the addresses of the NHIS households could not be disclosed to other contractors. I thought this restriction was incompatible with the development of a robust integrated survey design program, so prior to redesigning the NHIS based on the 1980 Decennial Population Census, I proposed using an area sampling method that did not depend on Census listings. The Bureau was perfectly willing to use area sampling instead of Census listings but said that it would be substantially more expensive—something like an additional million dollars.

I proposed the integrated sample design program and the NHIS sample design change to Dorothy Rice and she approved it. The next problem was how to get the additional million dollars needed for the NHIS redesign. The money was not included in the existing or future NCHS fiscal year budgets, and as I recall, the budget requests had already been transmitted to the Assistant Secretary for Health. With administrative help from Marjorie Greenberg, then working in the Office of the Assistant Secretary for Health, the justification for the amended budget request went forward and Congress appropriated the additional funds.

**Levy:** How did you implement the integrated survey design?

**Sirken:** Beginning in 1985, the first year that the area-based NHIS redesign was implemented, large pilot surveys were conducted using NHIS household and person listings as sampling frames for the NMES and NSFG. In these surveys, minority households were oversampled. The pilot studies demonstrated that integrated survey design was feasible and more efficient than the independently designed NMES and NSFG, and subsequently both surveys were linked to the NHIS. Currently, the Medical Expenditure Panel Survey, the NMES successor, conducted by Agency for Health Care Research and Quality, continues to use NHIS person and household listings as the sampling frames.

**Levy:** Integrated survey design research ultimately led you back to network sampling. How did that come about?

**Sirken:** In the early 1990s, a Panel of the Committee on National Statistics (CNSTAT) reviewed NCHS's plans for redesigning the health care provider surveys with samples of medical providers embedded in a subset of the NHIS psu's as proposed by the integrated survey design. Though the Panel liked the idea of linking sample designs of the health care provider surveys to the NHIS, it favored linkage at the health care provider level rather than at the psu level. In essence, the Panel proposed substituting the listings of health care providers reported by NHIS households for the complete sampling frames that listed all providers then in use by the health care provider surveys. With the nation's health care delivery system undergoing rapid change, the Panel felt that the NHIS-generated sampling frames would be easier to construct and maintain than the complete sampling frames.

In response to the Panel's proposal, NCHS initiated a research project that investigated the feasibility and efficiency of the Population Based Establishment Survey (PBES)—an establishment survey that uses a population survey-generated sampling frame. One of our first tasks was to derive expressions of a two-stage PBES unbiased estimator and variance of the volume of transactions between households and establishments. The problem eluded solution when the PBES was modeled as an establishment sample survey, but was easily solved subsequently when the PBES was modeled as a network sample household survey using a multiplicity rule that links an establishment's transactions with all households to every household with which the establishment has transactions (Sirken, Shimizu and Judkins, 1995). The PBES estimator and variance may be viewed as the two-stage cluster sampling adaptation of the multiplicity estimator and variance originally proposed by Dr. Birnbaum and myself.

## COGNITIVE ASPECTS OF SURVEY METHODOLOGY

**Willis:** What is CASM?

**Sirken:** CASM is the acronym of the Cognitive Aspects of Survey Methodology. CASM research is interdisciplinary and applies the theories and methods of the cognitive sciences in survey research.

**Willis:** Why would you, a mathematical statistician in a federal statistical agency, become interested in CASM?

**Sirken:** There was always a concern about the response error effects of NCHS's survey questionnaires, and considerable effort went into designing questionnaires to make them as accurate and as respondent-friendly as possible. My Office of Research and Methodology (ORM) reviewed the questionnaires designed by NCHS staffs, and it was a difficult task because questionnaire design is a craft rather than a science. I was ready and eager to embrace any scientific discipline that could provide a more scientific basis for survey questionnaire design.

**Willis:** Do you recall the time and circumstances when you first became interested in cognition and survey research?

**Sirken:** About 1980, I attended a two-day workshop convened by the Bureau of Justice Statistics that discussed the potential uses of cognitive methods to improve reporting of victimization by household respondents in the National Crime Survey. The workshop introduced me to a recent paradigm change in psychology with the two-stage stimulus/response model being replaced by a three-stage stimulus/cognition/response model with the focus on cognition and how the mind works during the cognitive stage. I got involved in the CASM movement in the early 1980s when I participated in the Advanced Research Seminar on the Cognitive Aspects of Survey Methodology, later known as the CASM I Seminar.

**Willis:** What was the CASM I Seminar and what was your role in it?

**Sirken:** The CASM I Seminar was sponsored by CNSTAT and supported by a National Science Foundation (NSF) grant. It was organized by Judy Tanur based on a proposal developed by Stephen Fienberg and Miron Straf. Tom Jabine was a consultant to the CNSTAT project. About 25 invited cognitive scientists and statisticians attended a six-day meeting in the summer of 1983 on the eastern shore of Maryland and a two-day follow-up session the following January in Baltimore. Questionnaires used in several national surveys were discussed at the Seminar including those of NCHS surveys, particularly NHIS. The Seminar was essentially the start of the CASM movement to foster interdisciplinary research that benefited survey research and cognitive psychology. An important factor in the Seminar's success was the availability of NSF funding to support meritorious CASM research projects that were developed at the Seminar. At the close of the June meeting, participants were encouraged to develop CASM research proposals based on the ideas they had presented at the Seminar.

When the CASM I Seminar reconvened, Robert Fuchsberg, NHIS Director, and I proposed a research project that compared two methods of designing the questionnaires for the forthcoming NHIS supplement on dental health. Using the traditional testing method, the dental health questionnaire would be tested under normal interviewing conditions in which Census enumerators interviewed NHIS respondents at their households. Using the proposed cognitive method, the dental health questionnaire would be tested in a laboratory setting in which professionally trained staffs conducted cognitive interviews with recruited respondents.

**Willis:** Was your research proposal funded by NSF?

**Sirken:** Initially, I was reluctant to submit our proposal to NSF because ORM staff and I were already overcommitted. However, I happened to discuss the proposal over lunch with Judy Lessler at the Joint Statistical Meetings in 1983 or '84, I believe, and she was very much interested. So after I got assurances that, if awarded, the NSF grant could cover her expenses and salary, and Judy got permission for a leave of absence from Research Triangle Institute, NCHS submitted the CASM proposal and was awarded the NSF grant. The following summer Judy moved to Washington, DC, and during the next year or so she, in collaboration with Roger Torangeau and Bill Salter, directed the CASM experiment. Findings of that study were published as the first report in the newly established series of NCHS reports on Cognition and Survey Measurement.

Findings of the CASM experiment were unexpected. They indicated that conventional and cognitive methods of testing the NHIS questionnaires were complementary rather than competitive as Bob Fuchsberg and I had originally anticipated. The testing methods exposed different rather than the same kinds of response problems. I recall meeting with Trish Royston and Debbie Trunzo, the ORM staff that reviewed NCHS questionnaires, and discussing implications of the experiment's findings for their work in reviewing NCHS questionnaires. We decided that the experimental findings implied that survey questionnaires should be laboratory tested before they are field tested. I concluded that a permanent Questionnaire Design Laboratory should be established to develop NCHS questionnaires. That's how the idea for a questionnaire design research laboratory was born.

ORM's proposal to establish a permanent Questionnaire Design Research Laboratory (QDRL) was approved by the NCHS Director, Dr. Manning Feinlieb, and ORM was given additional space for the Laboratory. The QDRL was established in 1985 with Trish Royston and Debbie Trunzo as co-directors. They deserve a lot of credit for the success of the QDRL. Testing and developing the questionnaires of the NHIS supplements, which changed annually, became the centerpiece of the Laboratory's work schedule. Laboratory methods were very successful in detecting and eliminating the kinds of glitches in survey questionnaires that are often missed in field testing. Soon requests to test the questionnaires of other federal agencies far exceeded the Laboratory's capacity and larger quarters and staff were allotted. News of the success of NCHS's QDRL spread quickly to other federal statistical agencies, and soon the Bureau of the Census and the Bureau of Labor Statistics established their own laboratories. Currently, cognitive testing of survey questionnaires is the common practice and several federal statistical agencies, here and abroad, have in-house questionnaire testing facilities.

Later, the NCHS, in collaboration with NSF, established the National Laboratory for Collaborative Research in Cognition and Survey Measurement (Sirken, 1991). It had two components. The QDRL was one component. It tested and designed the data collection instruments of federal surveys and was funded by NCHS and by the reimbursable work the Laboratory did for other agencies. Linda Pickle and Doug Hermann did pioneering work in the QDRL on the cognitive aspects of designing statistical maps. The second component, the Collaborative Research Program, was funded by the NSF, and fostered basic research on cognitive issues germane to improving data collection instruments of federal surveys. It included a Contract Research Program headed by Jared Jobe and a Visiting Scientist Program. About ten CASM research projects were completed before the Collaborative Research Program expired in the early 1990s. Findings of these research projects appear in NCHS Series 6 reports on "Cognition and Survey Measurement."

**Willis:** I recall working with you and Gad Nathan on a project funded by the Visiting Scientist Program.

**Sirken:** Yes, Gad was a Visiting Scientist from the Hebrew University and the project was "Cognitive Aspects of Designing Sensitive Survey Questions." Our experiments compared cognitive and behavioral theories of the likelihood of truthful response (LTS) in surveys on illicit drug use. We did not actually recruit drug users in the Laboratory, but asked laboratory subjects to respond to survey questionnaires as if they were the drug users depicted in vignettes. In brief, the cognitive theory posited that the LTS depends on the drug users' perceptions of the disclosure risks of truthful response and the consequences, and the behavioral theory posited that the LTS depends on two survey features, the type of illicit drug (marijuana or cocaine) and the extent of privacy protection (confidentiality or anonymity) the survey provided. The comparisons slightly favored the cognitive theory.

**Willis:** So overall did the Laboratory work out as you had hoped?

**Sirken:** Yes and no. The influence and success of the QDRL in applying cognitive techniques to design survey questionnaires far exceeded my expectations. On the other hand, I regret the demise of the Collaborative Research Program in NCHS and wish it could be revived. It was great while it lasted.

**Willis:** There was also the CASM II Seminar in 1996. How did that develop and what were its objectives?

**Sirken:** About 1993, the tenth anniversary of the CASM I Seminar, I proposed convening a CASM II Seminar to expand the scope of CASM research beyond questionnaire design to other survey design features, and to expand the scope of interdisciplinary relationships beyond cognitive psychology to other disciplines. The CASM II Seminar was jointly funded by NSF and NCHS. It was organized by a planning committee of interdisciplinary-minded survey researchers, and was managed and supported by ORM staff, especially Susan Schechter and Karen Whitaker and Tom Jabine, an ORM consultant. The six-day Seminar met in Charlottesville, Virginia during June, 1996 and featured 15 commissioned papers by survey methodologists and cognitive scientists, which along with remarks of discussants appear in the book published by Wiley (Sirken et al., 1999).

Like the CASM I Seminar, the CASM II Seminar sought to develop interdisciplinary research project proposals. Unfortunately, funding was unavailable to support CASM II research proposals. Even before the CASM II Seminar convened, I was beating the bushes looking for funds.

**Willis:** And that's how you got involved in the Funding Opportunity in Survey Research?

## THE FUNDING OPPORTUNITY IN SURVEY RESEARCH

**Sirken:** A potential source of funding for CASM II research proposals surfaced toward the end of the CASM Seminar II when Cheryl Eavey, Head of NSF's Methodology, Measurement and Statistics program, offered to administer and fund a grants program in basic survey research for a three-year period if a consortium of federal statistical agencies provided matching funds. After some false starts, I took NSF's proposal to the Federal Committee on Statistical Methodology (FCSM), an interagency committee of federal statisticians to which I belong, and requested its help in recruiting a consortium of federal statistical agencies to put up the matching funds.

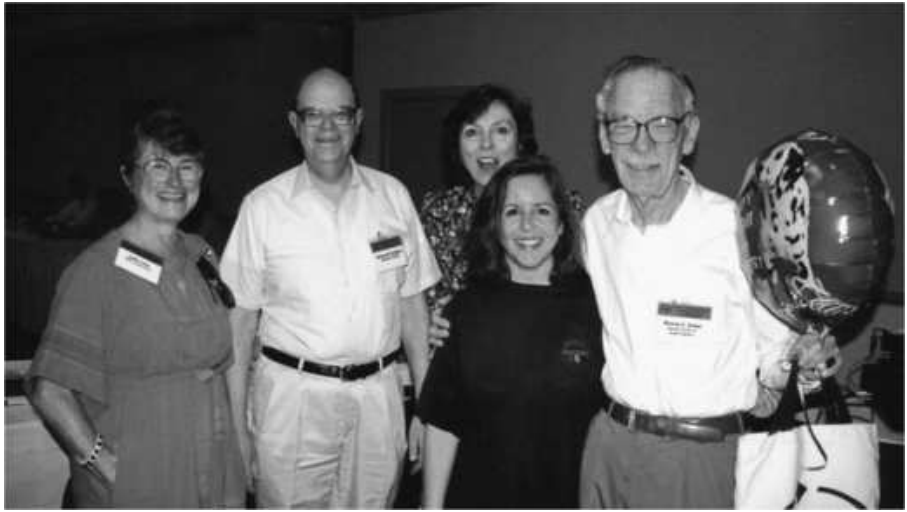

Fig. 5. *CASM II Seminar, 1996. From l. to r.: Judy Tanner, Norman Bradburn, Barbara Wilson, Susan Schechter, Monroe Sirken.*

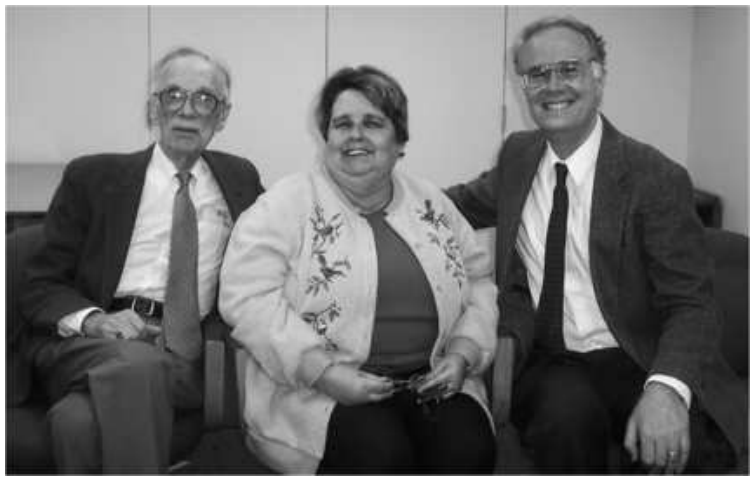

Fig. 6. *NCHS Questionnaire Design Research Laboratory, 2006. From l. to r.: Monroe Sirken, Mary Moien, Barry Graubard.*

**Willis:** How did FCSM react to the NSF proposal?

**Sirken:** FCSM appointed a research subcommittee to draft a proposal to the Interagency Committee on Statistical Policy (ICSP), a committee of the directors of about 15 of the largest federal statistical agencies. Thirteen ICSP agencies agreed to individually support the Funding Opportunity during a three-year period and in January, 1999, the NSF issued the first announcement of the Funding Opportunity in Survey Research, a grants program in interdisciplinary survey and statistical research that is oriented to the future methodological needs of federal statistical agencies. In 2001, the ICSP agencies and the MMS extended their agreement beyond 2002.

**Willis:** Is the Funding Opportunity working out as you hoped?

**Sirken:** During its initial seven years, 1999–2006, the Funding Opportunity made grants totaling more than four million dollars and supported about 25 investigator-initiated research projects. It is too early to assess the long-term contributions of these projects, but I'm optimistic. Arguably the most outstanding early achievement of the Funding Opportunity has been the successful development of the infrastructure of a complex grants program that is oriented to long-term needs of the federal statistical system, and is funded jointly by a consortium of federal statistical agencies and the NSF and administered jointly by the FCSM research committee and the NSF. The Funding Opportunity provides a vehicle for statistical agencies in our decentralized federal statistical system to collaborate in funding methodological research which none of them could individually afford. It provides the FCSM with a vehicle to carry out its basic mission of improving the quality of federal statistics. Yes, I would say, the Funding Opportunity has worked better than I ever imagined in fostering basic survey research that is oriented to the needs of federal statistical agencies.

## RECENT YEARS: 1996–

**Willis:** I think we have pretty well covered your major activities until the late 1990s. What have you been doing lately?

**Sirken:** I have been a Senior Research Scientist at NCHS during the past ten years. I spend full time conducting and fostering survey and statistical research as I no longer have administrative or policy responsibilities. I divide my time between conducting statistical research mostly on network sampling, often in collaboration with Iris Shimizu (Sirken and Shimizu, 2007), and fostering interdisciplinary survey research, mostly as chair of the FCSM Research Committee that steers, oversees and evaluates activities of the Funding Opportunity in Survey and Statistical Research (Sirken, 2004). I find the work of conducting and fostering scientific research very satisfying and highly recommend that combination of activities to senior scientists.

## ACKNOWLEDGMENTS

Many thanks to Mary Moien for organizing and coordinating the conversation, to Karen Whitaker for making arrangements to conduct the conversation in the NCHS Questionnaire Design Research Laboratory, and to Debbie Trunzo, Ed Korn and Tom Fears for editorial comments. Opinions expressed in this conversation are Monroe's and do not necessarily represent the views of the National Center for Health Statistics.